\documentclass[conference]{IEEEtran}
\usepackage{cite}
\usepackage{amsmath,amssymb,amsfonts}
\usepackage{algorithmic}
\usepackage{graphicx}
\usepackage{textcomp}
\usepackage{xcolor}
\def\BibTeX{{\rm B\kern-.05em{\sc i\kern-.025em b}\kern-.08em
    T\kern-.1667em\lower.7ex\hbox{E}\kern-.125emX}}

\usepackage{xspace} 
\usepackage{comment} 
\usepackage{catchfile} 
\usepackage{multicol} 
\usepackage{multirow} 
\usepackage{enumitem} 

\usepackage{fancyvrb} 
\usepackage{fvextra}  
\usepackage{listings} 
\usepackage{wrapfig}
\usepackage{tcolorbox}

\usepackage[hyphens]{url} 
\usepackage{hyperref}
\hypersetup{breaklinks=true}

\definecolor{light-grey}{gray}{0.95}

\usepackage{subcaption} 
\usepackage{catchfile} 

\newcommand{\jh}[1]{}

\IEEEoverridecommandlockouts
\IEEEpubid{\begin{tabular}[t]{@{}l@{}}
\copyright{}2022 IEEE. Personal use of this material is permitted. Permission from \\
IEEE must be obtained for all other uses, in any current or future media,\\
including reprinting/republishing this material for advertising or promotio- \\
nal purposes, creating new collective works, for resale or redistribution to\\
servers or lists, or reuse of any copyrighted component of this work in\\ 
other works.\hfill\end{tabular}
\hspace{\columnsep}\makebox[\columnwidth]{ }}

\begin{document}
%
%


\title{Towards Automated PKI Trust Transfer for IoT}


\author{\IEEEauthorblockN{Joel~H\"oglund, Shahid~Raza}
\IEEEauthorblockA{RISE Research Institutes of Sweden\\
Isafjordsgatan 22, 16440 Kista, Stockholm \\
\{joel.hoglund, shahid.raza\}@ri.se}
\and
\IEEEauthorblockN{Martin~Furuhed}
\IEEEauthorblockA{Technology Nexus Secured Business Solutions, Sweden.\\Telefonv\"agen 26, 12626 H\"agersten, Stockholm \\
martin.furuhed@nexusgroup.com}
}

\maketitle

\begin{abstract}
IoT deployments grow in numbers and size and questions of long time support and maintainability become increasingly important. To prevent vendor lock-in, standard compliant capabilities to transfer control of IoT devices between service providers must be offered. 
We propose a lightweight protocol for transfer of control, and we show that the overhead for the involved IoT devices is small and the overall required manual overhead is minimal. We analyse the fulfilment of the security requirements to verify that the stipulated requirements are satisfied.


\end{abstract}

\begin{IEEEkeywords}
security, IoT, PKI, digital certificates, enrollment, embedded systems
\end{IEEEkeywords}



\section{Introduction}
\label{sec:intro}

The increasing number of IoT devices used worldwide for safety and security critical applications such as grid infrastructure and e-health highlights the need for robust and scalable security solutions suitable for IoT. The last couple of years have seen an increase in protocols and standards targeting the Internet of Things, including standards covering security aspects. These standards define security services such as relatively lightweight secure communication and authentication. Together with recent proposals for key establishment and certificate enrollment, important steps towards bringing Public Key Infratructure, PKI, to IoT have been taken, towards making IoT devices first class Internet citizens. 

Our ultimate goal is a complete and automated PKI that scales to billions of IoT devices. Before the goal has been reached, a number of issues remain before IoT developers and providers have access to PKI solutions for all their security needs.

With an increase in the number of IoT deployments, questions of long time support and maintainability become increasingly important. Among the open issues are how to handle the transfer of trust, when the responsibilities of maintenance of IoT devices are shifted from one service provider to another. The scenario cannot be captured with a single protocol, but needs to be mapped out and covered by references to existing solutions together with new proposals where there currently are gaps. A specific goal with the work presented here is to provide a clear guide for how it can be done with minimal overhead in terms of manual labour. This leads to the following
problem formulation: what is the minimal procedure needed, in terms om manual intervention, to securely shift the operation of one IoT device from one service provider to another?

The criteria for a complete and successful transfer of trust is in terms of when all involved IoT devices have enrolled and received new operational certificates, making them recognized as valid participants of the target organization PKI, while meeting all the requirements defined for the proposed protocol.


The main contributions of the paper are as follows:
\begin{itemize}
\item A design of a lightweight schema for trust transfer, which allows control of IoT deployments to shift between service providers in a highly automated manner.
\item A feasibility study using a prototype implementation for constrained IoT devices. 
\item An security analysis to show that the schema meets the stated requirements.
\end{itemize}

The rest of this paper is organized as follows:
Section \ref{sec:background} presents a brief discussion of vital concepts for the proposed protocol.
Section \ref{sec:related} presents related work.
Section \ref{sec:threat} gives our threat model and assumptions. 
Section \ref{sec:requirements} formalizes the requirements of the proposed protocol. 
Section \ref{sec:trusttransfer} presents a detailed scenario together with our proposal for formalizing the steps into a protocol with a maximal level of automation.
Section \ref{sec:evaluate} presents the results of the feasibility evaluation. In section \ref{sec:security_assesment} we present the assessment of the security requirements, before concluding the paper.

\begin{figure}[th]
\centering
  \begin{subfigure}{0.24\textwidth}
    \includegraphics[width=\linewidth]{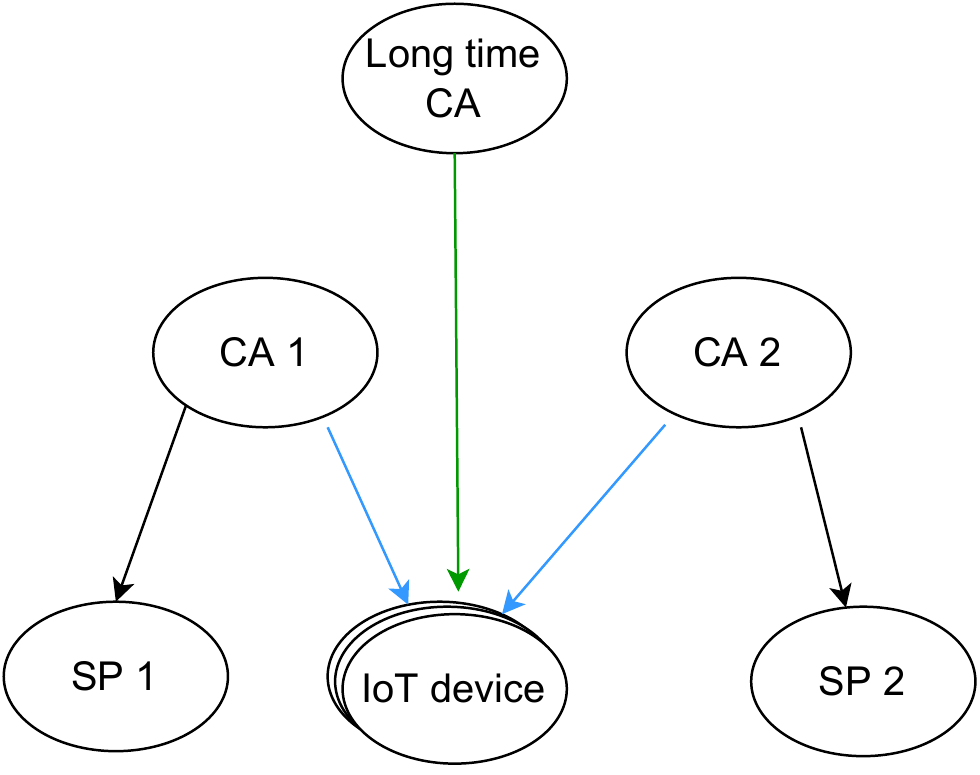}
    \caption{}
    \label{fig:CA_hierarkies_a}
  \end{subfigure}%
  \hspace*{\fill}   
  \begin{subfigure}{0.24\textwidth}
    \includegraphics[width=\linewidth]{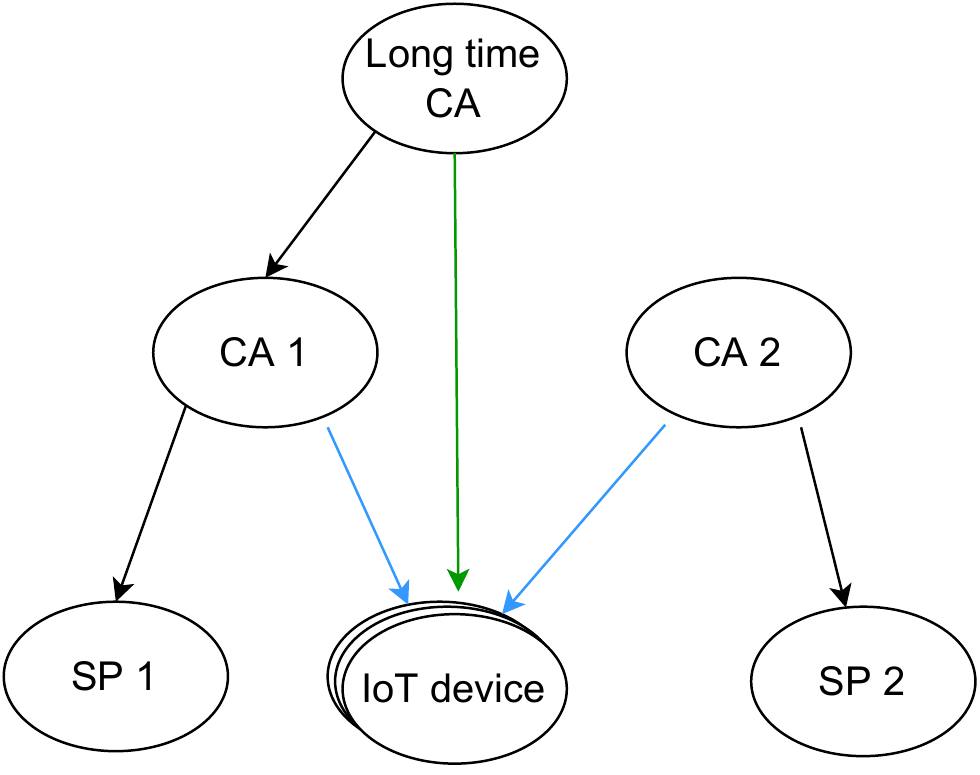}
    \caption{}
    \label{fig:CA_hierarkies_b}
  \end{subfigure}%
  \\

  \hspace*{\fill}   
  \begin{subfigure}{0.24\textwidth}
    \includegraphics[width=\linewidth]{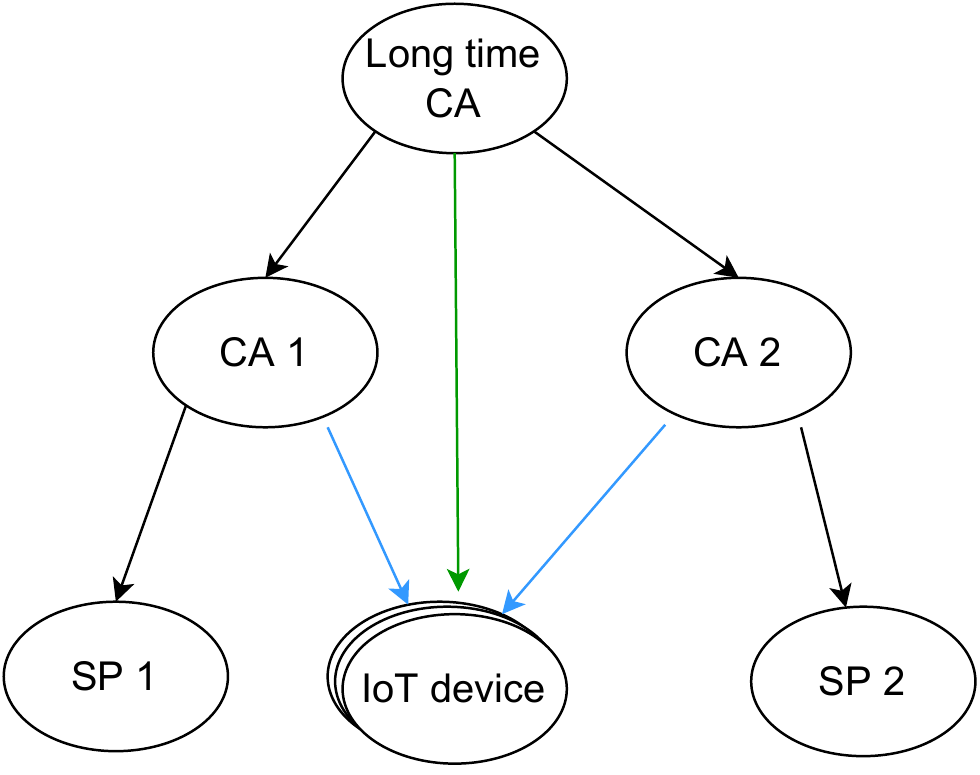}
    \caption{}
    \label{fig:CA_hierarkies_c}
  \end{subfigure}%
  \hspace*{\fill}   
  \begin{subfigure}{0.22\textwidth}
    \includegraphics[width=\linewidth]{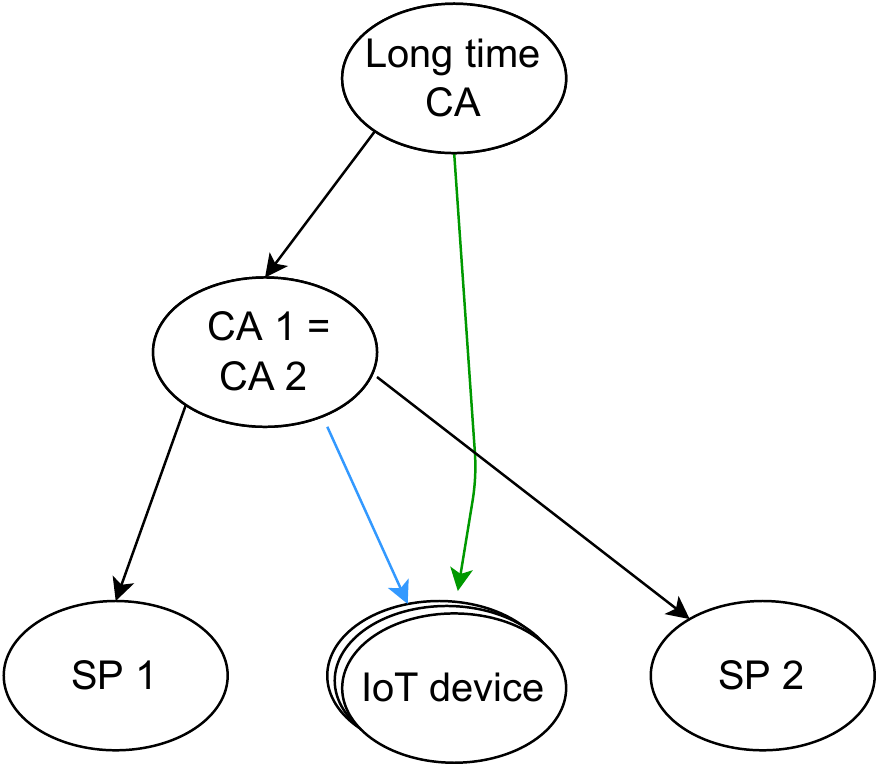}
    \caption{}
    \label{fig:CA_hierarkies_d}
  \end{subfigure}
\caption{Different options for CA hierarchies. All arrows represent certificate issuing, green arrows for factory certificates, blue arrows for operational IoT certificates} \label{fig:CA_hierarkies}
\end{figure}

\section{Towards Automated PKI: needed mechanisms and concepts}
\label{sec:background}
This section introduces mechanisms related to the creation of a PKI for IoT and concepts needed for the rest of the paper.

\subsection{Security services and PKI}
Two of the key security services needed to implement a system which can establish and maintain trust from the system perspective are authorization and authentication. 

An authorization mechanism ensures that an actor can perform exactly the actions they are entitled to and no other actions. To build an authorization service, a secure authentication service is a needed building block: authorization of actions requires authenticating the actor. The role of the authentication service is to provide the necessary trustworthy binding between an entity and a public key. 

The full system needed to manage the authentication services and their artefacts, certificates, keys, policies and roles forms a Public Key Infrastructure, PKI.

\subsection{PKI hierarchies and Trust Transfer within IoT}
PKIs rely on authentication through publicly available keys encapsulated in certificates which are signed by a certificate authority, CA. The CA is in turn identified by its certificate, either self-signed or signed by yet another CA, forming a hierarchy up to a self-signed top/root CA. The system allows chains of certificates to be verified up to the top nodes, which need to be already trusted \cite{PKI-RFC2459}. To bootstrap these trust chains, the party performing the authentication must have trusted access to the self signed root node certificates. For an IoT device this means it must be equipped with the necessary root certificates in a trust store, either through factory pre-programming or through enrollment operations\footnote{In the last couple of years PKI enrollment solutions suitable also for IoT have been shown feasible~\cite{pki4iot, LICE}.}.

At the lowest layer of the CA hierarchies are the IoT devices, and the servers belonging to service providers with which the devices need to communicate. Different possible CA hierarchies which involve IoT devices and two different service providers are illustrated in Fig. \ref{fig:CA_hierarkies}. The task of transferring trust for an IoT deployment becomes equivalent with securely updating the device membership from the original PKI to the target PKI. \jh{Minimal definition. What do you think?}

For trust transfer scenarios the implications of the different CA hierarchy types are the following: 
If the trust hierarchies are completely separated, as in Fig. \ref{fig:CA_hierarkies_a}, the IoT device needs to be equipped with a root certificate for CA1 in advance of the first enrollment, to be able to authenticate CA1. Similarly the device needs to be updated with a root certificate for CA2 in advance of the trust transfer.
If the CA1 is a sub-CA of the permanent CA, as in Fig. \ref{fig:CA_hierarkies_b}, it is sufficient to provide an update with the a root certificate for CA2. For the scenarios in Fig. \ref{fig:CA_hierarkies_c} and \ref{fig:CA_hierarkies_d}, all entities can be authenticated using only prior access to the certificate of the permanent CA. These are the minimal requirements for which root certificates that must be added to the IoT device trust store. For performance reasons, additional certificates can be added to later enable authentication through certificate references, in which case it is sufficient for the communicating parties to only send hashes of certificates. This type of reference based public key usage is supported in EDHOC based key establishment \cite{EDHOC}.

\section{Related Work}
\label{sec:related}

\textit{Ownership transfer}
The related area of IoT ownership transfer has been studied from different perspectives. In \cite{Khan1392237_chownIoT} the focus is on the privacy and protection of smart home device data. A custom solution for creating user profiles, and automatically detecting ownership changes for individual devices is presented. Compared with our efforts, this is on the opposite end standard compliance, where automatization is used not for reducing costs and handling scale, but for convenience of individual users and end user privacy protection.

In \cite{SecureOwnershipTransfer_RISE} a custom non-standard solution is proposed, where the authors specifically do not assume PKI support from the devices. Their focus is on ensuring forward and backward security between the former and new owner. The solution is based on symmetric keys and a trusted third party. Despite the differences in assumptions concerning PKI support and standard compliance, they investigate a similar scenario as we do, and some of their requirements have relevance for our solution as well.

\section{System and threat model}
\label{sec:threat}
We consider primarily IoT deployments where multiple devices communicate with a limited number of Internet servers. The IoT devices are constrained in terms of both bandwidth and computational resources. They are capable of asymmetric crypto operations, but energy constraints might make it important to keep them at a minimum. The deployed devices are often communicating using wireless low power networks, in which the packet sizes are severely restricted and packet losses are common.

We rely on the Dolev-Yao adversarial model \cite{1056650_DOLEV}, which assumes that an attacker can eavesdrop any message being sent, can record messages and inject both old messages and modified ones into the communication. On the other hand we assume that IoT devices themselves are not being tampered with, and the adversary cannot break crypto functions within the relevant time span.

We assume that the involved service providers establish mutual trust, in such a way that they will not actively attack the counterpart. They might still be interested in gathering leaked data, unless prevented. We present ways to lessen the assumption of mutual honesty through remote attestation.


\section{Requirements} %
\label{sec:requirements}
Based on the above description of challenges and threats we arrive at the following trust transfer protocol requirements.

\paragraph{FR1: Impersonation security} The protocol must be capable of preventing an adversary from impersonating either a legitimate IoT device, or any of the involved parties.

\paragraph{FR2: Freshness / replay attack resistance} Eavesdropping traffic and replaying captured messages must not break the security guarantees.

\paragraph{FR3: Forward security} The old service provider shall not get access to any private data which can compromise the privacy of the new service provider and its onwards operations.

\paragraph{FR4: Backward security} The new service provider shall not get access to any private data belonging to the old service provider, which is not explicitly agreed to be shared.

These functional requirements are in common with the ones proposed in \cite{SecureOwnershipTransfer_RISE}. In addition we identify the following non-functional requirements:

\paragraph{NFR1: Automatization} The protocol must offer the desired functionality with a minimum of manual intervention. 

\paragraph{NFR2: Resource efficiency} The protocol must allow all operations directly involving the IoT devices to be lightweight to run on relatively resource constrained devices.

\paragraph{NFR2: Standard compliance} To be a feasible for adoption by industry, the protocol must build upon existing and ongoing standardization efforts wherever possible.

\begin{figure}[t!]
\centering
\includegraphics[width=0.22\textwidth]{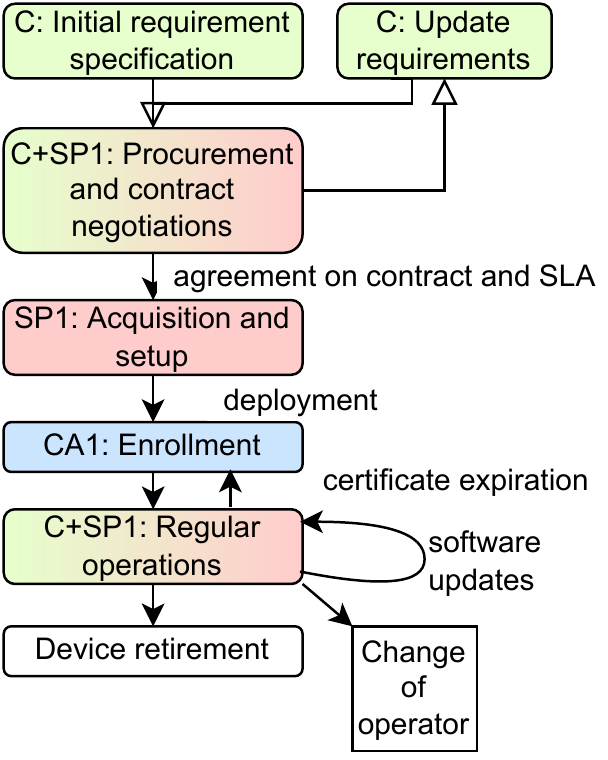}
\caption{\label{fig:lifecycle}The IoT life cycle, involving customer, service provider and certificate authorities}
\end{figure}

\begin{figure}[t!]
\centering
\includegraphics[width=0.35\textwidth]{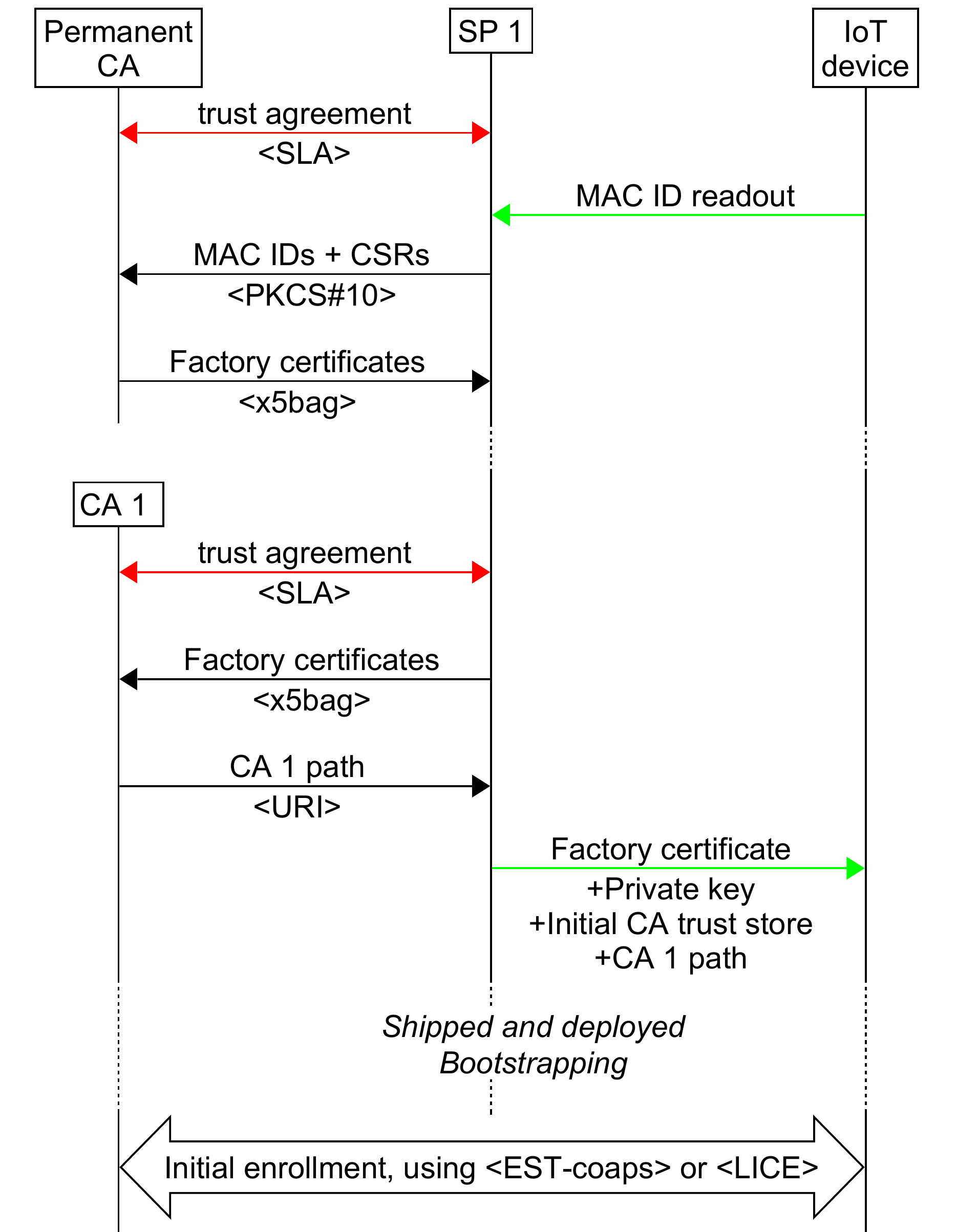}
\caption{\label{fig:trusttransfer_enroll}The IoT device initial life cycle stages, with standards used for setup and enrollment (see \ref{sec:device}). Red arrows correspond to operations where manual intervention is expected. Green are setup and deployment specific, while black are standard based and fully automated.}
\end{figure}

\section{IoT life cycle}
For an IoT device to be a part of a PKI constitutes the main enabler for a number of crucial security services, and is absolutely necessary for the goal of offering standard based interoperability and preventing vendor lock in. 
Making IoT devices parts of a PKI is a nontrivial task for constrained devices. To give the context for how the task can be achieved, we present existing and proposed solutions for how an appropriate environment for trust transfer can be created. We cover the first stages in the PKI for IoT life cycle, while adhering to existing standards for all steps wherever possible. A high level overview of the life cycle is shown in Fig. \ref{fig:lifecycle}. A more detailed diagram of the initial life cycle phases is given in Fig. \ref{fig:trusttransfer_enroll}.

\paragraph{Scope and limitations}
We address issues directly related to PKI management, needed to guarantee the required security services. A deployment might have additional functional requirements such as downtime constraints which need to be factored in when scheduling the actions to be performed.

\paragraph{Involved actors}
\label{involved_actors}
In the first steps of the life cycle description the following actors and roles are relevant to clarify:

{\bf CA}: A reliable, well established, certificate authority (Permanent CA): A certificate authority which can be trusted for an extended period of time, suitable for providing trust root(s) to the device's initial truststore.

{\bf SU}: The IoT service user, who is also the system owner (owner/customer). This is the actor; company or organization, who uses the IoT system to achieve a goal. The goal can be internal, with the SU as the service end user, or as part of providing a service to others.

{\bf SP1}: The initial IoT service provider; the company which is in charge of configuring the IoT devices, installing and maintaining them.

{\bf CA1}: The initial operational CA; the certificate authority with which SP1 has made an agreement to provide operational certificates, including renewals when needed. It can be the same as the permanent CA.

\subsection{Procurement and SLAs}
\label{sec:contract}
The starting point for the scenario is that a company or an organization, SU, has identified a need which can be fulfilled with an IoT system. The IoT system needs to be clearly specified, ordered, deployed and thereafter maintained. The deployment could be within the SU's own premises, or within any other area where they have obligations to perform monitoring or offer services which can be aided by the IoT installation. 

As part of the procurement process the SU specifies service level agreement (SLA) conditions that must be met. In this work we focus on the directly PKI related conditions. This includes to specify that the chosen IoT service provider must be able to transfer the role of system maintainer to a new service provider without breaching agreed security guarantees. The demands could also specify additional criteria for minimal service disruptions during any system update.

In line with the efforts to lessen the burden of manual intervention in any software service operation, SLAs can be used to formalize contractual agreements in a manner suitable for automated checking \cite{8591028_DSLA_SC_B}. From the perspective of our trust transfer proposal, details on how SLAs are monitored and acted upon are outside the scope.

An IoT provider who accepts the required conditions gets the order. Together the SU and the IoT service provider, hereafter SP1, formalize the requirements in a contract containing the agreed upon SLA. Besides quality of service specifications, the parties clarify the service endpoints to be used for accessing services and data.

\subsection{Device acquisition, factory credential and firmware preparations}
\label{sec:device}
The SP1 acquires IoT devices which meet the customer's functional requirements, as well as non-functional requirements in terms of security protocol support and update capabilities. The section corresponds to the Acquisition and setup-stage of Fig. \ref{fig:lifecycle}.

A crucial part of an automated PKI capable of handling IoT devices without manual intervention is how to prepare the devices such that they can perform initial authentication operations once deployed. The practical solution for mutual authentication is to pre-program devices with a secret factory key and a factory certificate, plus an initial truststore containing server certificates. The device needs the server certificates forming the certificate chain up to the CA root of the factory certificate plus, based on communication needs, root certificates to authenticate also servers with certificates belonging to other root CAs.

All IoT devices come with unique IDs when they are delivered from the manufacturer. In the following we assume that the SP1 uses these unique device IDs as the basis for the device names in the factory certificates.
The IDs might be simply matched between a list of IDs and a sticker on the device, or through a QR code, or extracted through some programming port. The exact measures will depend on the device type at hand. 

If the IoT device is equipped with a secure and protected module, it could implement the 802.1AR standard for Secure Device Identities, DevIDs \cite{802.1AR-DevID}. The hardware requirements make the standard less suitable for the most constrained IoT devices, but for sufficiently capable devices the module can be used to offer protection also from physical tampering.

The SP1 has an agreement with a CA which they trust, to order long lived factory certificates. This agreement must match the conditions in the SLA with the SU regarding predictable long time availability of the CA. Since the factory certificates should have a lifetime corresponding to the lifetime of the IoT device it is extra important that there, with a high likelihood, will be an entity available which can reply to inquires about the certificate revocation status for all of the expected device lifetime. The factory certificates should be restricted in terms of operational capabilities. The initial post-deployment enrollment is what assigns an operational certificate to the device, with the needed capabilities to operate within the SP1 infrastructure.

The SP1 generates cryptographic keypairs and creates certificate signing requests, CSRs, for all IoT devices that should receive factory certificates. The requests are communicated to the permanent CA, which creates factory certificates and sends them back. This communication takes place over the regular Internet, and is not restricted in terms of bandwidth. The certificate signing requests can be sent using the PKCS\#10 standard~\cite{RFC2986-PKCS10}. Since the targets are IoT devices the certificates should be compact. The proposed C509 standard \cite{I-D.ietf-cose-cbor-encoded-cert} offers a more efficient format compared with X509, using ECC cryptography for the strongest cryptographic guarantees at relatively short key lengths. The CSRs as well as the replies can be sent one by one as needed, or collected and sent in batches. All of the communication happens over a TLS secured communication link.

SP1 contacts a CA which will act as the operational CA, CA1. Unless CA1 is the same as the permanent CA, CA1 needs to be updated about the identities of the devices for which it should grant operational certificates to. This is solved by sharing the factory certificates. A proposed format with minimal overhead is x5bag, in which certificates are wrapped in byte strings and placed in a CBOR array~\cite{I-D.ietf-cose-x509}. In return, the SP1 is given the URI which the IoT devices should contact for doing the enrollment of operational certificates.

The data exchange between the SP1 and the CA1 can be fully automatised, given a pre-existing contract which specifies the rights for any device which can authenticate itself using a private key corresponding to one of the shared factory certificates to request an operational certificate.

At this point the SP1 is equipped with the data needed to do the initial programming of devices, which provides the devices with the initial firmware, including factory private key, factory certificate, initial truststore and information of the CA-URI. The initial programming and data transfer to the IoT devices takes place in a trusted environment.

The steps covered until this point are illustrated in Fig. \ref{fig:trusttransfer_enroll} up until "Shipped and deployed".

\subsection{Deployment and initial enrollment}
The device is physically installed in its target environment. This can be done by SP1, by the SU or by a trusted third party. In the following we assume that deployment specific bootstrapping issues have been solved.

Upon startup the IoT device contacts the CA1 to do initial enrollment for an operational certificate. The device authenticates itself through the factory certificate which is registered with the CA1. This certificate also serves to authorize the certificate request. The mutual authentication is done as part of establishing a secure channel, using either a DTLS or EDHOC handshake.

After the mutual authentication the IoT device sends a certificate signing request to the operational CA, using the proposed C509 CBOR format \cite{I-D.ietf-cose-cbor-encoded-cert}, or the less compact PKCS\#10-format for legacy systems. The CA replies with an operational certificate, in either X509 or C509 format. The choice of format depends on whether the enrollment is done following EST-coaps \cite{EST-RFC9148} or the proposed more compact EDHOC based enrollment protocol~\cite{LICE}.

IoT devices with sufficient computational resources are capable of generating the key-pair themselves, which is the preferred solution whenever available, as the private key never needs to leave the device. For the most constrained devices the enrollment is done with the inclusion of a server generated key-pair.

\subsection{Normal operations}
After enrollment the IoT device is equipped with an operational certificate which is recognized by the servers it needs to communicate with, and has an updated truststore which ensures that the device can perform authentication of all endpoints of relevance.

During the normal operations the SP1 ensures the IoT devices are kept up to date with software upgrades, following the SUIT architecture mechanisms~\cite{RFC9019_SUIT}. Before the operational certificate expires the device will do re-enrollment with CA1.

\section{IoT Trust Transfer}
\label{sec:trusttransfer}

\subsection{Introduction and problem formulation}
In general terms, the IoT service user, SU, decides that they want to switch service provider for their IoT services, while maintaining their existing deployments and installations. This is the high level goal which should be achieved with a minimum of service disruptions and minimal need of human intervention. Today the operations needed for a secure transfer of control between service providers is insufficiently specified. Without clear protocols, the task becomes at the best very labour intensive, with several manual steps which needs to be tailor-made to the specific scenario. At worst, impossible. 

\begin{figure}[t!]
\centering
\includegraphics[width=0.33\textwidth]{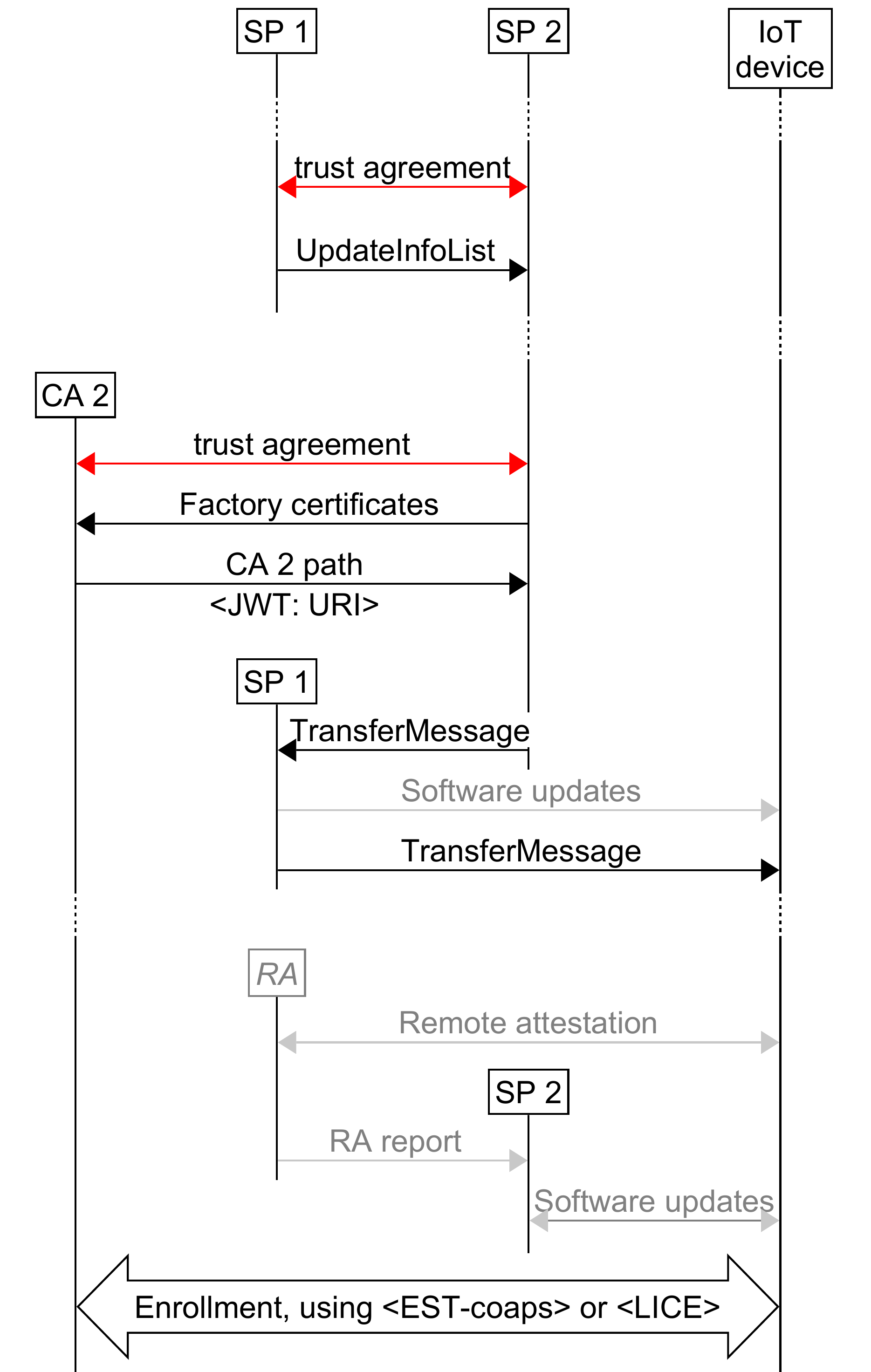}
\caption{Operator change. Automated operations in black, optional operations in gray}
\label{fig:updateprotocol}
\end{figure}

In the following we detail the steps, referring to existing standards where applicable, and proposing solutions for missing parts. An illustration of the protocol flow is given in Fig. \ref{fig:updateprotocol}, which will be referred to in the following subsections.

\subsection{Additional involved actors}
In addition to the actors introduced in \ref{involved_actors}, the following are included. \textbf{SP2}: A second IoT service provider; the company selected by the SU to overtake the responsibilities to maintain the IoT devices from SP1. \textbf{CA2}: second operational CA; the certificate authority with which SP2 has made an agreement to provide operational certificates.

\subsection{Preparations for operator change}

If the need arises for the customer to switch service providers, the initial contract (see \ref{sec:contract}) specifies that the current service provider SP1 needs to contact the designated new service provider, SP2. This step might include manual efforts, in forming a specific contract which specifies the details of transactions which are about to take place. Specifically, it needs to specify a starting date from when SP2 must be ready to start maintaining the IoT devices, within the total allowed time-span defined by the SU. 

SP1 and SP2 need to agree on the state of the IoT firmware, in particular which services and which versions of the services the IoT devices will provide at the time of shifting the maintenance responsibilities. A solution to automatize the auditing of the IoT device state is to use remote attestation.

\textit{Remote attestation}, RA, is an advanced security service that has attracted considerable attention the last couple of years. In remote attestation a device produces a proof of its current state, which is checked and verified by a trusted third party to be in accordance with the expected output.

To offer strong security guarantees RA relies on access to a trusted hardware component for the device being attested, such as TPM or ARM TrustZone. More constrained IoT devices do not have access to these dedicated hardware resources. There are also software based RA solutions, and hybrid versions with limited requirements on protected memory areas. There is active research in the area \cite{s21051598_remote_attestation_survey} as well as large ongoing IETF standardisation efforts \cite{I-D.ietf-rats-architecture}.

In addition to agreeing on RA details, the parties declare which certificates that are to be used for signing of protocol data. When the trust relationship is established and a transfer specification contract is formed, the old service provider can share device information with the new service provider. The information exchange needs to contain the following data items:

- The factory certificates for every involved IoT device for which the responsibility of maintenance is about to be transferred from SP1 to SP2. 

- The earliest and the latest switch-over time for each involved device.

- Firmware code and/or service description(s) of the software that the IoT device is running. There are several possible alternatives, depending on if SP2 is to continue using the same software that is already available, and to what degree the source code of the components is shared. We propose the state of the device software is shared through sharing references to the relevant SUIT manifests. 

- Optionally, if RA is to be performed, SP1 needs to share the information needed for a verifier to evaluate the response from the device being attested. 

The mandatory information represented as a CBOR array is specified in CDDL as follows:

\begin{lstlisting}[caption=UpdateInfoList,label=label:uil, breaklines=true]
UpdateInfoList = [* DeviceUpdateInfo]

DeviceUpdateInfo = (
  factoryCertificate:   TBSCertificate,
  updateTimeNotBefore:  Time,
  updateTimeNotAfter:   Time,
  versionInfo:          (suit-manifest-seq-number,
                        suit-reference-uri),
)
\end{lstlisting}
This update information, encoded as an array of pairs, is signed by SP1 using JSON Web Signatures and the previously agreed identity.

The designated SP2 needs to perform the same procedure with an CA of choice that SP1 carried out together with CA1 before initial deployment: forward the factory certificate list to CA2 and get a CA-URI token back. In addition to these administrative steps the SP2 configures a update server endpoint, and prepares a TransferMessage, following the format given below.


\begin{lstlisting}[caption=TransferMessage,label=label:transfermessage]
TransferMessage = (
  ResetTimeNotBefore:  Time,
  ResetTimeNotAfter:   Time,
  raURI:               bstr / null,
  updateURI:           (bstr, bool),
  enrollURI:           bstr,
  fallbackURI:         bstr   
)
\end{lstlisting}

If remote attestation is used, the TransferMessage contains the RA URI. The updateURI is set to the own update server, with a flag to indicate if devices should contact the update server before the enrollment. Finally the CA2 path is set as the enrollURI. This set of claims is treated as the payload of a COSE\_Sign1 object, which is signed by the SP2 key agreed upon in the first phase of the protocol, resulting in a signed CBOR Web Token.

\textit{Protocol bindings:} LwM2M is a possible device management protocol which could be used for this purpose, together with SUIT. 

\subsection{Performing the service provider change}
When SP1 has received the TransferMessage from SP2, it validates the signature, parses the set of claims and copies the fields, except the fallback URI which is set to the SP1 update server, into individual messages for each target IoT device. SP1 can if needed perform a last remote software update to the target devices. The resulting CWTs are signed by SP1 and sent to target IoT devices. After the TransferMessage has been received and validated, the individual IoT devices reset to a state agreed upon in the agreement between SP1 and SP2, where the resulting state includes updated information about which new server endpoints to contact.

Upon restarting, the device will optionally contact the RA server to participate in a RA challenge response. Thereafter, depending on the updateURI flag, it can contact the SP2 update server. Finally the device does re-enrollment with CA2, using a certificate signing request in either CBOR or PKCS\#10-format. The device will receive a new operational certificate, recognized by the relevant SP2 endpoints, as well as additional needed truststore updates. Just like in the initial enrollment situation, the IoT device trusts the new CA after mutual authentication.

It should be noted that the device truststore after the last SP1 operation must contain certificates capable of authenticating CA2. Additionally if RA is used, or pre-enrollment SP2 updates are needed, the trust roots of the RA server and the SP2 update server endpoint must be present in the trust store. The least complex scenario is when the SP2 endpoint can be authenticated by certificates in the IoT truststore in its initial state. This is trivially the case when the CA hierarchies correspond to \ref{fig:CA_hierarkies_c} or \ref{fig:CA_hierarkies_d}. Otherwise there must be a truststore update operation which is not rolled back by the SP1 reset operation. 

If any of the steps permanently fails, such as RA failure or failure to authenticate with the CA2 or the SP2 update server, the IoT device will use the fallback URI to once more contact the SP1 update server. For completeness, SP1 might now require the device to perform a new RA session, to verify its state after interactions with SP2.

\subsection{Continued operations and certificate revocation checking}
After the new enrollment operations the device is fully reconfigured as part of the SP2 management domain, and will communicate with the SP2 servers based on its new configuration. In the proposed protocol the effort to check the revocation status of IoT device certificates, both operational and the long term factory certificates, is put on the Internet servers. They can handle existing relatively heavy-weight protocols such as OCSP or CRLs. To extend revocation checking capabilities to constrained devices, more efficient mechanisms are needed.

\section{Feasibility study}
\label{sec:evaluate}
Based on the protocol design goals, to target resource constrained devices, it is critical to show that the protocol overhead is sufficiently small to match expected IoT capabilities. In the following we validate the proposed building blocks in terms of messaging, computational and memory overhead.
Our tests have been performed on the nRF52840-DK platform, which is a relatively powerful but relevant target IoT device with an Arm Cortex-M4, 802.15.4-radio and 256 kB RAM\footnote{www.nordicsemi.com/Products/Development-hardware/nRF52840-DK}.

\subsection{Messaging overhead}
To evaluate the feasibility of the protocol, and the overhead for IoT devices we calculate the sizes of involved messages and transactions.

As can be seen in table \ref{table:message_sizes} the TransferMessage, the protocol message specifically sent to the IoT devices, constitutes only a few hundred bytes. Since this is small compared with the handshake and enrollment operations,  networks and devices which are capable of handling the related PKI operations will have no difficulties with the added protocol messages.

\newcommand{\mytable}{\input{protocol_message_sizes.table}}
\begin{table}[ht]
\caption{Protocol message size in bytes}
\centering
\resizebox{0.4\textwidth}{!}{%
\begin{tabular}{|l c c|} 
  \input{protocol_message_sizes.table}
\end{tabular}
}

\label{table:message_sizes}
\end{table}

\subsection{Computational overhead}
With the exception of the remote attestation operations, which are highly dependent on the type of RA performed, the only added operation with significant computational impact for the IoT devices is the signature checking of the TransferMessage. The signature checking of the COSE\_Sign1 is the same type of operation which is performed as part of an EDHOC handshake. On the nRF52840 platform, a relevant target IoT device with an Arm Cortex-M4, the signature verification operation takes 21 ms, when the signature is done using the commonly used P-256 curve. This can be compared with a full EDHOC handshake which needs around 90 ms of active CPU time when using the same ECC curve.

\subsection{Memory overhead}
The functionality needed for the IoT authentication operations are of the same type that are used for EDHOC and OSCORE. By reusing the crypto liberaries, no extra memory footprint will be taken into account for crypto operations, and only a few hundred bytes for the TransferMessage specific handling. Our implementations of required crypto functionality used by both OSCORE and EDHOC needs approximately 6~KB of ROM, plus 5~KB more of EDHOC specific code, for the nRF52840 platform.

Solutions for remote attestation of IoT devices have been successfully emulated on IoT devices as limited as the old TmoteSky platform with 48 kB ROM, 10 kB of RAM and access to 1MB of flash \cite{RA_SARA}, and could therefor coexist with the required PKI components on more capable devices such as nRF52840.

\subsection{Non-functional requirement compliance}

The security requirements are assessed below in \ref{sec:security_assesment}. Here we focus on evaluating the compliance with the non-functional requirements.

\paragraph{NFR1: Automatization}
\label{sec:economical}
The feasibility analysis illustrates that besides the initial trust agreements and SLA establishments, all other operations can be fully automated. This is a key requirement to enable large scale IoT deployments with PKI support, through the reduction of the PKI costs per device.

Currently the pricing models for CA services are complex, and dependent on a long range of customer requirements. The requirements can be both security guarantees, such as requirements on dedicated hardware security modules (HSM) and organizational constraints, such as which of the organizational constructs depicted in Fig. \ref{fig:CA_hierarkies} that need to be supported\footnote{Nexus company policies}~.

Specifically for the cost of individual certificates, for the few CA providers which share any certificate pricing information openly online, the lowest per certificate cost found is starting from 7.95 USD per year, as of April 2022 \cite{ComodoSSLstore}. This price range is infeasible for large scale IoT deployments.

The current situation illustrates the need for a continued development towards standards, increased automatization and reduced costs per device.

\paragraph{NFR2: Resource efficiency} All the needed building blocks have been demonstrated in versions suitable for modern constrained IoT devices. Since the transfer functionality is vital, but rarely used, it is crucial to reuse already existing crypto functionality on the device, resulting in a minimal added overhead.

\paragraph{NFR3: Standard compliance} All security critical components are contained within existing or proposed standards. The combination of secure upgrades and remote attestation is still an area where only initial standardisation solutions have been proposed. The modular approach proposed for our trust transfer solution makes it relatively easy to upgrade parts of the protocol to incorporate for example new RA mechanisms, or new crypto algorithms to be used for authentication or encryption services.

\section{Security Assessment}
\label{sec:security_assesment}
The security assessment of the protocol builds upon the derivations done in the SIGMA paper~\cite{SIGMA}. A correctly constructed protocol will keep the security properties offered by the individual components, and hence be capable of offering the intended security services as long as the components keep their security guarantees.

\subsection{Security requirement compliance}
For each of the requirements listed in \ref{sec:requirements} we analyse how the claim is supported by the protocol. 

\paragraph{FR1: Impersonation security} All protocol participants have well defined credentials which they use for authentication. All protocol interactions happen over authenticated sessions. As long as the credentials used for authentication, and the crypto mechanisms used for encryption/decryption are not compromised, no internal or external party will be able to take on a role they have not been given.

The IoT device is a special case. If it keeps the key-pair corresponding to the old SP1 operational certificate, it could impersonate its own old role after the transfer operation. To prevent this, SP1 should instruct CA1 to revoke the operational certificates. The revocation will cover interactions with servers and endpoints already capable of handling CRLs. To also cover cases when the IoT device communicates with other resource constrained devices, an OCSP solution for IoT is needed.

\paragraph{FR2: Freshness} For the secure sessions, either a combination of TLS+DTLS or TLS+OSCORE are used. They all include information based on sequence numbers in the messages, allowing the opposite endpoints to detect and discard replayed messages. This protects against outside eavesdroppers, who capture and replay full packages. The mechanisms are not designed to  prevent an endpoint from resending the same application layer content multiple times.

\paragraph{FR3: Forward secrecy} After the re-enrollment with CA2, IoT devices will protect their onward communication using a new key-pair. Unless the SP1 has installed and left a secret backdoor available at the device, SP1 will no longer have access to any new data related to the device or the new service provider. For the SP2 to ensure this is not the case, a remote attestation schema could be employed. 
\paragraph{FR4: Backward secrecy} After the IoT reset operations, the device will only contain the information that SP1 has agreed to share with SP2. It is the responsibility of SP1 to not grant access to any sensitive information based on either the new operational CA2 issued certificate, or the long lived factory certificate that will remain.

\section{Conclusion}
When IoT deployments become more common and grow in size, issues of long time maintenance and the scalability of the security services become critical. Making use of proposed and available PKI solutions suitable for IoT we have proposed a lightweight protocol for the transfer of control of IoT deployments, with a minimal manual overhead. The solution ensures the possibility of long time support for IoT deployments, preventing vendor lock in. We have shown that given the integrity of the secure building blocks, the protocol maintains the desired security properties.

\section*{Acknowledgment}
This research is partially funded by the Swedish SSF Institute PhD grant and by the EU H2020 projects ARCADIAN-IoT (Grant ID. 101020259) and CONCORDIA (Grant ID: 830927).

\bibliography{ref}
\bibliographystyle{IEEEtran}

\end{document}